\documentclass[aps,prab,reprint,floatfix,superscriptaddress,amsmath,amssymb]{revtex4-2}

\usepackage{graphicx}
\usepackage{bm}
\usepackage{booktabs}
\usepackage{tikz}
\usetikzlibrary{arrows.meta,decorations.pathmorphing}
\usepackage[colorlinks=true,linkcolor=blue,citecolor=blue,urlcolor=blue]{hyperref}

\newcommand{\dd}{\mathrm{d}}

\begin{document}

\title{GEM-PIC-RZ: a Galilean electromagnetic particle-in-cell code\\ in quasi-cylindrical geometry with nonuniform grids}

\author{Alexander Pukhov}
\affiliation{Institut f\"ur Theoretische Physik I, Heinrich-Heine-Universit\"at D\"usseldorf, 40225 D\"usseldorf, Germany}

\date{\today}

\begin{abstract}
Laser-plasma accelerators span scales from the micrometre wavelength of the driving laser to acceleration lengths of millimetres or metres. Full particle-in-cell (PIC) simulations resolve the laser carrier but are constrained by the Courant condition, whereas quasi-static models are computationally efficient but do not resolve it. We present GEM-PIC-RZ, a quasi-cylindrical implementation of the Galilean electromagnetic PIC scheme. The method retains the slow evolution of the forward field and therefore resolves the laser carrier while allowing a propagation step governed by the evolution of the laser and plasma rather than by the cell size. User-defined nonuniform grids and a charge-conserving regularization of the bubble closure support efficient simulations across these scales. Comparisons with the full PIC code FBPIC show agreement for the laser wake, the depletion and redshift of a few-cycle pulse, and ionization injection; the measured CPU time is about 2\,\% of that for FBPIC in the reported runs;  the injected charge carries a model uncertainty of about 15\,\% from the regularization. An envelope model, in contrast, underestimates the depletion of the few-cycle pulse fivefold. Self-injection in a sharp density down-ramp is not yet reproduced and remains an open problem.
\end{abstract}

\maketitle

%---------------------------------------------------------------------------
\section{Introduction}\label{sec:intro}

Plasma-based accelerators driven by lasers~\cite{Tajima1979,Esarey2009} or particle beams sustain fields roughly three orders of magnitude stronger than radio-frequency cavities, and their design rests on simulations. These simulations are multiscale: the laser wavelength $\lambda_0 \sim 1\,\mu$m, the plasma wavelength $\lambda_p \sim 10$--$100\,\mu$m and the acceleration length of millimetres to metres span six orders of magnitude. Full electromagnetic PIC codes resolve $\lambda_0$ and are bound by the Courant condition, so the number of time steps is the acceleration length divided by a fraction of $\lambda_0$. Quasi-cylindrical (azimuthal-mode) codes such as Calder-Circ and FBPIC~\cite{Lifschitz2009,Lehe2016} reduce the cost per step, and the Lorentz-boosted frame~\cite{Vay2007} and Galilean spectral solvers~\cite{Lehe2016b} reduce the number of steps or suppress the numerical Cherenkov instability, but the Courant limit remains.

Quasi-static codes~\cite{Sprangle1990,Mora1997,Lotov2003,Huang2006,Mehrling2014,Li2021,Diederichs2022,FerranPousa2019} remove it by dropping \emph{all} derivatives with respect to the slow time. The plasma response of a slice then follows from the sources of that slice, and the driver is advanced with a step set by its own evolution. The price is the laser carrier: lasers enter through ponderomotive envelope models~\cite{Mora1997,Benedetti2018}, and ionization in the instantaneous field, few-cycle pulses with strong redshift, harmonics and the radiation of the accelerated beam are outside the model.

The Galilean electromagnetic particle-in-cell (GEM-PIC) algorithm~\cite{GEMPIC} closes this gap. It transforms Maxwell's equations and the Vlasov equation to the Galilean coordinates $\zeta = z - ct$ and a slow coordinate $s$, either $s = ct$ (GT1) or $s = z$ (GT2), and splits the transverse field into forward and backward components. The slow derivative is kept for the forward component, which carries the laser, so the carrier is resolved and propagates, diffracts, depletes and redshifts self-consistently. It is dropped only for the backward and longitudinal components, where it appears together with the much larger $\zeta$-derivative. The scheme is therefore \emph{not} a quasi-static approximation, yet it has no Courant limit, because the field update is semi-implicit. In three dimensions GEM-PIC reproduced a 31.8\,cm laser wakefield accelerator with ionization injection simulated with FBPIC~\cite{GEMPIC}.

Many laser and particle-beam drivers are approximately axisymmetric, so a small number of azimuthal modes can represent them at substantially lower cost than a three-dimensional simulation. Here we present GEM-PIC-RZ, an implementation of GEM-PIC in quasi-cylindrical geometry, and assess its performance against full PIC for three central laser-plasma-accelerator problems: nonlinear wake formation, laser evolution over multiple Rayleigh lengths, and electron injection. Section~\ref{sec:model} gives the reduced equations in azimuthal modes and the particle scheme. Section~\ref{sec:disc} describes the discretization on nonuniform radial grids, its dispersion relation, and two numerical problems with their remedies: a numerical Cherenkov radiation behind relativistic bunches and the excitation of spurious modes by the singular bubble closure. Section~\ref{sec:impl} summarizes the implementation. Section~\ref{sec:valid} verifies the code against analytic references, and Sec.~\ref{sec:lpa} compares it with FBPIC for four laser-plasma problems, three of which it reproduces and one, density down-ramp injection, which it does not yet. Section~\ref{sec:fel} outlines the extension to free-electron lasers.

%---------------------------------------------------------------------------
\section{Model}\label{sec:model}

This section states the equations that GEM-PIC-RZ solves: the reduced Maxwell equations of GEM-PIC in azimuthal modes and the motion of the particles in the Galilean variables. The derivation in Cartesian geometry is given in Ref.~\cite{GEMPIC}; here we show how it carries over to cylindrical geometry and which choices the code makes.

\subsection{Variables and field equations}

We use units with $c = 1$, lengths in $1/k_0$, time in $1/\omega_0$ ($k_0 = \omega_0/c$ the laser wavenumber), densities in the critical density $n_c$, momenta in $m_e c$ and fields in $m_e c\,\omega_0/e$. Following Ref.~\cite{GEMPIC}, the fast coordinate is $\zeta = z - t$, and the slow coordinate is either $s = t$ (GT1, $\partial_t = \partial_s - \partial_\zeta$, $\partial_z = \partial_\zeta$) or $s = z$ (GT2, $\partial_t = -\partial_\zeta$, $\partial_z = \partial_s + \partial_\zeta$). The reduced field equations derived below are the same for both; the particles of GEM-PIC-RZ are advanced in GT2 (Sec.~\ref{sec:particles}), so a level $s$ of the simulation is the plane $z = s$ of the laboratory, and the slice $\zeta$ of that level is the time $t = s - \zeta$ at which the fields pass the plane. The simulation window covers $0 \le \zeta \le L_w$; its front is at $\zeta = L_w$. Every field is expanded in azimuthal modes,
\begin{equation}
F(r,\varphi,\zeta,s) = F_0 + 2\,\mathrm{Re}\sum_{m=1}^{M} F_m(r,\zeta,s)\, e^{im\varphi},
\end{equation}
and we drop the mode index below. A linearly or circularly polarized laser lives in $m = 1$; $m = 0$ carries the wake.

The transverse fields are combined into the characteristic components
\begin{align}
R_r &= E_r + B_\varphi, & L_r &= E_r - B_\varphi, \nonumber\\
R_\varphi &= E_\varphi - B_r, & L_\varphi &= E_\varphi + B_r .
\end{align}
A wave propagating in the $+z$ direction at the speed of light has $L = 0$ and depends only on $\zeta$; $R$ is the forward (laser) field. With GT1, Maxwell's equations become, exactly,
\begin{align}
\partial_s R_r &= \partial_r E_z + \frac{im}{r} B_z - j_r, \label{eq:Rr}\\
\partial_s R_\varphi &= \frac{im}{r} E_z - \partial_r B_z - j_\varphi, \label{eq:Rp}\\
(\partial_s - 2\partial_\zeta) L_r &= -\Bigl(\partial_r E_z - \frac{im}{r}B_z + j_r\Bigr), \label{eq:Lr}\\
(\partial_s - 2\partial_\zeta) L_\varphi &= -\Bigl(\partial_r B_z + \frac{im}{r}E_z + j_\varphi\Bigr), \label{eq:Lp}\\
(\partial_s - \partial_\zeta) B_z &= -\frac{1}{r}\partial_r (r E_\varphi) + \frac{im}{r} E_r, \label{eq:Bz}\\
(\partial_s - \partial_\zeta) E_z &= \frac{1}{r}\partial_r (r B_\varphi) - \frac{im}{r} B_r - j_z . \label{eq:Ez}
\end{align}
The forward field varies fast in $\zeta$ (on $\lambda_0$) and slowly in $s$ (on the Rayleigh length and the depletion length). The backward field $L$ and the longitudinal fields $E_z$, $B_z$ of a co-propagating structure vary on the same $\zeta$ scales as their sources, which move with $v \approx c$, so their $s$-derivatives are small compared with their $\zeta$-derivatives. With GT2 the same equations hold with $\partial_s - 2\partial_\zeta \to -\partial_s - 2\partial_\zeta$ in Eqs.~\eqref{eq:Lr}, \eqref{eq:Lp} and without $\partial_s$ in Eqs.~\eqref{eq:Bz}, \eqref{eq:Ez}~\cite{GEMPIC}. Neglecting $\partial_s$ wherever it appears together with $\partial_\zeta$ gives, for both transformations, the reduced system
\begin{align}
\partial_\zeta L_r &= \tfrac12\Bigl(\partial_r E_z - \frac{im}{r}B_z + j_r\Bigr), \label{eq:Lr2}\\
\partial_\zeta L_\varphi &= \tfrac12\Bigl(\partial_r B_z + \frac{im}{r}E_z + j_\varphi\Bigr), \label{eq:Lp2}\\
\partial_\zeta B_z &= \frac{1}{r}\partial_r (r E_\varphi) - \frac{im}{r} E_r, \label{eq:Bz2}\\
\partial_\zeta E_z &= -\frac{1}{r}\partial_r (r B_\varphi) + \frac{im}{r} B_r + j_z , \label{eq:Ez2}
\end{align}
together with Eqs.~\eqref{eq:Rr}, \eqref{eq:Rp}, where $E = (R + L)/2$, $B_\varphi = (R_r - L_r)/2$ and $B_r = (L_\varphi - R_\varphi)/2$. These are the azimuthal-mode form of the universal system, Eqs.~(49)--(54) of Ref.~\cite{GEMPIC}: with $(x, y) \to (r, \varphi)$, $ik_x \to \partial_r$ (or $r^{-1}\partial_r r$ in the curls) and $ik_y \to im/r$, the equations agree term by term. Equations \eqref{eq:Lr2}--\eqref{eq:Ez2} are integrated from the front of the window, where $L = E_z = B_z = 0$, towards the rear. In vacuum the system gives $\omega = 2\kappa k_\perp^2/(4\kappa^2 - k_\perp^2) \simeq k_\perp^2/(2\kappa)$ for a mode $\propto e^{i\kappa\zeta - i\omega s}$, the paraxial limit of the exact $\omega = \sqrt{\kappa^2 + k_\perp^2} - \kappa$; in a cold plasma $k_\perp^2$ is replaced by $k_\perp^2 + \omega_p^2$~\cite{GEMPIC}. The backward wave is not described; Ref.~\cite{GEMPIC} shows that the forward branch of the full dispersion relation approaches that of the reduced system for $\kappa \gg \omega_p$.

\subsection{Particles}\label{sec:particles}

The particles follow the characteristics of the Vlasov equation in GT2~\cite{GEMPIC},
\begin{equation}
\frac{\dd s}{\dd\zeta} = \frac{v_z}{v_z - 1}, \qquad \frac{\dd \bm r_\perp}{\dd\zeta} = \frac{\bm v_\perp}{v_z - 1}, \qquad \frac{\dd \bm p}{\dd\zeta} = \frac{\bm F}{v_z - 1}, \label{eq:char}
\end{equation}
with the Lorentz force $\bm F$ of the fields reconstructed from the modes. A plasma particle at rest keeps its $s = z$ and crosses the whole window within one level, as in the lab frame. A particle whose $s$ reaches the next level $s^1$ before it leaves the window stays inside it: it is trapped. In principle all particles can be integrated along $\zeta$; as recommended in Ref.~\cite{GEMPIC}, GEM-PIC-RZ integrates the fast, trapped ones along $s$ instead, which is explicit and more accurate. Each particle is thus in one of two lists.

\emph{Streaming particles}: the plasma electrons, mobile ions and electrons freed by ionization. A fresh layer of plasma is loaded at the front of the window at every step $s^0 \to s^1 = s^0 + \Delta$, with the density $n(s^1)$ and a quiet start in $r^2$ and $\varphi$; one macroparticle represents the plasma of a ring of axial thickness $\Delta$. The layer then streams through the whole window, using $\zeta$ as the independent variable, Eq.~\eqref{eq:char}. Between two slices with the spacing $h_n$ a particle is pushed over the time
\begin{equation}
\dd t = \frac{h_n}{1 - v_z}, \label{eq:dt}
\end{equation}
by a two-pass Boris push~\cite{Boris1970} that makes $\dd t$ consistent with the mean $v_z$ over the interval; its lab position $z = s$ advances by $v_z\,\dd t$. It deposits its current with the weight $W = q\,\dd t/\Delta$, the charge that crosses the cell during one step [Eq.~(99) of Ref.~\cite{GEMPIC}]. The fields are those of the level $s^1$ being computed (Sec.~\ref{sec:march}).

\emph{Beam particles}: bunches and trapped particles, which move with the window. They are advanced in $s$ from $s^0$ to $s^1$ ($\dd t = \dd s/v_z$, $\dd\zeta = (v_z - 1)\dd t$) with the Vay pusher~\cite{Vay2008} in sub-steps of at most $\Delta z_\mathrm{max}$, in the fields of level $s^0$; the Vay pusher keeps the cancellation $\bm E + \bm v\times\bm B \approx 0$ of a relativistic beam in its own field at any step, which the Boris pusher does not. A beam particle deposits a current with the weight $q/v_z$, linear in $\zeta$.

The two lists exchange particles. A streaming particle whose lab position reaches $z = s^1$ inside the window has been trapped; it joins the beam list if $v_z \ge 0.9$, where the push along $\zeta$ becomes stiff [$\dd\zeta/\dd t = v_z - 1 \to 0$]. A slower one, which may happen in a strong laser field where electrons move forward by more than $\Delta$ while crossing the pulse, continues to stream in the next step from its $\zeta$. Conversely, a beam particle that becomes slower than $v_z = 0.8$ returns to streaming. No particle is deleted, so the charge is conserved.

\subsection{Initial fields, ionization and ions}

A laser pulse is initialized in mode $m=1$ as a forward wave, $R_{r,1} = P^*$, $R_{\varphi,1} = iP^*$, $L = 0$, with $P = E_x + iE_y$ a paraxial Gaussian beam (with wavefront curvature and Gouy phase at a distance from focus) or a given radial profile.

A bunch of velocity $\beta$ starts with its own co-moving field, $-(\nabla_\perp^2 + \gamma^{-2}\partial_\zeta^2)\psi = j_z/\beta$, $R = (1 + \beta)\bm E_\perp$, $L = (1-\beta)\bm E_\perp$, $E_z = -\gamma^{-2}\partial_\zeta\psi$, with $\bm E_\perp = -\nabla_\perp\psi$. The problem is solved with the radial eigenvectors of the discrete Laplacian of the field solver and one tridiagonal solve along $\zeta$ per eigenvector, so a bunch in vacuum is an exact discrete equilibrium. A field that is not the co-moving one would stay behind a bunch with $v < c$ that slips back in the window and exchange energy with it; this matters for the FEL application (Sec.~\ref{sec:fel}).

Ion species can be mobile, with their own density profile (e.g.\ a dopant section), and field-ionizable. The ADK tunnel rate~\cite{ADK1986,Chen2013} of each level is evaluated with the instantaneous $|\bm E|$, including the laser carrier, and averaged over the two slices of the interval; ionization energies are taken from the NIST database~\cite{NIST}. A Monte Carlo step with exponential waiting times allows several levels per interval. The random numbers come from a counter-based hash of (seed, particle, step, slice, pass), so all back-ends draw the same numbers.

%---------------------------------------------------------------------------
\section{Discretization}\label{sec:disc}

The discretization has to meet three requirements. It must keep the constraints of Maxwell's equations on a radial grid that is fine at the axis and coarse outside, since the plasma response and the laser are concentrated near the axis while the fields extend far out. It must keep each slice cheap, since a slice is solved once or twice per step for every one of thousands of slices. And it must be stable for steps $\Delta$ much larger than the cell size. This section shows how the scheme meets them and which numerical modes it leaves, and how they are controlled.

\subsection{Radial grid and operators}

The radial grid has faces $\rho_i$, $i = 0,\dots,N$ ($\rho_0 = 0$ is the axis), and centres $c_i$, $i = 0,\dots,N-1$, $\rho_i < c_i < \rho_{i+1}$. Faces carry $R_r$, $L_r$, $B_z$, $j_r$; centres carry $R_\varphi$, $L_\varphi$, $E_z$, $j_\varphi$, $j_z$. With the finite-volume measures
\begin{equation}
A^f_i = \rho_i (c_i - c_{i-1}), \quad A^f_0 = \tfrac12 c_0^2, \qquad A^c_i = c_i(\rho_{i+1} - \rho_i),
\end{equation}
the operators are
\begin{align}
(G e)_i &= \frac{e_i - e_{i-1}}{c_i - c_{i-1}}, \qquad
(g b)_i = \frac{b_{i+1} - b_i}{\rho_{i+1} - \rho_i}, \\
(\mathrm{CE})_i &= \frac{c_i E_{\varphi,i} - c_{i-1}E_{\varphi,i-1}}{A^f_i} - \frac{im}{\rho_i}E_{r,i}, \\
(\mathrm{CB})_i &= \frac{\rho_{i+1}B_{\varphi,i+1} - \rho_i B_{\varphi,i}}{A^c_i} - \frac{im}{c_i}B_{r,i},
\end{align}
for $\partial_r$ to the faces and to the centres and for the two curls in Eqs.~\eqref{eq:Bz2}, \eqref{eq:Ez2}. Two discrete identities hold exactly on any grid,
\begin{equation}
\mathrm{CE}\bigl(G e,\ \tfrac{im}{c} e\bigr) = 0, \qquad \mathrm{CB}\bigl(\tfrac{im}{\rho} b,\ -g b\bigr) = 0, \label{eq:curlgrad}
\end{equation}
because $A^f_i = \rho_i(c_i - c_{i-1})$ and $A^c_i = c_i(\rho_{i+1}-\rho_i)$; they are the discrete forms of $\nabla\times\nabla = 0$ in the two staggerings. The divergence on the measures $A^c$ makes the discrete Gauss law exact. The particles deposit with area weighting plus a quadratic correction per interval, $w = x + \kappa x(1-x)$ with $x$ linear in $r^2$, where $\kappa$ is chosen so that the deposited volumes equal $A^f$, $A^c$: a uniform plasma is deposited exactly, also on a stretched grid. Fields are gathered linearly in $r$, which is exact near the axis for fields $\propto r$.

\subsection{The march in $\zeta$}\label{sec:march}

The window has slices $\zeta_0 = 0 < \zeta_1 < \dots < \zeta_{N_\zeta}$ with spacings $h_n = \zeta_{n+1} - \zeta_n$. Fields live on the slices, currents on the intervals (half-cells) between them. A step computes the level $s^1$ from $s^0$ by marching from the front slice to the rear. At slice $n$ the slice $n+1$ of level $s^1$ is known.

Equations~\eqref{eq:Rr}, \eqref{eq:Rp} are integrated with the trapezoidal rule in $s$,
\begin{equation}
R^1_n = R^0_n + \frac{\Delta}{2}\bigl[\mathcal F(E_z, B_z)^0_n + \mathcal F(E_z, B_z)^1_n\bigr] - \frac{\Delta}{4}\,\mathcal J_n, \label{eq:Rdisc}
\end{equation}
where $\mathcal F$ is the right-hand side without the current and $\mathcal J_n$ is the sum of the currents of the two adjacent half-cells at both levels. On a nonuniform $\zeta$ grid the current at the slice is the linear interpolation $(h_n j_{n-1/2} + h_{n-1} j_{n+1/2})/(h_{n-1} + h_n)$. Equations~\eqref{eq:Lr2}--\eqref{eq:Ez2} are integrated with the trapezoidal rule over the half-cell $h_n$, e.g.
\begin{equation}
L^1_n = L^1_{n+1} - \frac{h_n}{4}\bigl[\mathcal G^1_{n+1} + \mathcal G^1_n\bigr] - \frac{h_n}{2}\, j_{n+1/2}. \label{eq:Ldisc}
\end{equation}
$R^1_n$ and $L^1_n$ are linear in the unknown $E_z$ and $B_z$ of slice $n$. Inserted into the trapezoidal rules for Eqs.~\eqref{eq:Bz2}, \eqref{eq:Ez2}, the identities~\eqref{eq:curlgrad} remove $E_z$ from the $B_z$ equation and $B_z$ from the $E_z$ equation. What remains are two independent tridiagonal systems per mode,
\begin{align}
\bigl[16 - h_n(2\Delta + h_n)\,\nabla^2_{f,m}\bigr]\, B_{z,n} &= \text{known}, \\
\bigl[16 - h_n(2\Delta + h_n)\,\nabla^2_{c,m}\bigr]\, E_{z,n} &= \text{known} - 16 h_n\, j_{z,n+1/2},
\end{align}
with the discrete Laplacians on the faces and on the centres. $R$ and $L$ then follow explicitly. The matrices depend only on $m$, $\Delta$ and $h_n$; they are factorized once per distinct $h_n$ and recomputed only when $\Delta$ changes. The cost of a slice is linear in $N$ and $M$.

The currents of half-cell $n+\frac12$ at level $s^1$ depend on the fields of slice $n$, which are being computed. The march therefore uses a predictor--corrector: the predictor takes the fields of slice $n$ from the old level, $F^1_n \approx F^0_n$ (the slow evolution in $s$), the streaming particles are pushed through the half-cell and deposit, the slice is solved, and the push and solve are repeated with the new fields (two passes by default). When the half-cell is final, $R$ at slice $n+1$ is corrected for the difference between the predicted and the final current that entered its Eq.~\eqref{eq:Rdisc}. At the rear slice the half-cell behind the window is set equal to the last one (zero gradient); a zero current there would halve the source of $R$ and produce a growing spike. Near the outer radial boundary an optional absorber damps $R$.

\subsection{User-defined nonuniform grids}\label{sec:grids}

The physics of a plasma accelerator is concentrated in small regions: near the axis, at the laser pulse, at a witness bunch, in a density ramp. All three axes therefore use the same mapping: the step $h(x)$ is given at reference points and is linear in between, and the nodes follow from $\dd x/\dd k = h(x)$ with $k$ the cell or step index. Within a segment the steps grow geometrically and the nodes are analytic. All slice equations above hold for any $h_n$, and a uniform grid is the special case of a constant table and is reproduced bit for bit.

\subsection{Numerical dispersion}\label{sec:dispersion}

The step $\Delta$ is meant to be orders of magnitude larger than the cell size, so the stability of the scheme has to be shown, not assumed. For a single vacuum Fourier mode $\propto e^{ik_\zeta\zeta - i\omega s}$ with transverse wavenumber $k_\perp$ ($\nabla_\perp^2 \to -k_\perp^2$) on a uniform $\zeta$ grid, the scheme of Sec.~\ref{sec:march} gives
\begin{equation}
\tan\frac{\omega\Delta}{2} = \frac{\Delta\, h\, k_\perp^2 \sin\theta}{16\sin^2(\theta/2) - h^2 k_\perp^2\cos^2(\theta/2)}, \qquad \theta = k_\zeta h. \label{eq:disp}
\end{equation}
This is the relation of the 1D plasma model of Ref.~\cite{GEMPIC}, Eq.~(84) there, $\tan(\omega\Delta/2) = \omega_p^2\Delta h\sin\theta/[16\sin^2(\theta/2) - \omega_p^2h^2]$, with $\omega_p^2$ replaced by $k_\perp^2$; the factor $\cos^2(\theta/2)$ comes from the average of the transverse fields over the two slices of a half-cell in the RZ scheme. $\omega$ is real for all $k_\zeta$, $k_\perp$, $\Delta$ and $h$, so the scheme has no growing modes in vacuum, and there is no Courant condition linking $\Delta$ to $h$ or to the radial cells. With a plasma the phase $\omega\Delta/2$ passes $\pi/2$ where the denominator changes sign, which gives the practical limit $\Delta h \le 4/\omega_p^2$ of Ref.~\cite{GEMPIC}; in typical laser wakefield runs ($\omega_p^2 = 10^{-3}$, $h = 0.25$) it allows $\Delta$ of order $10^4$. For $\theta \to 0$ and $\Delta \to 0$, $\omega \to 2k_\zeta k_\perp^2/(4k_\zeta^2 - k_\perp^2) \simeq k_\perp^2/(2k_\zeta)$, the continuum relation of the reduced system and the paraxial diffraction of the exact $\omega = \sqrt{k_\zeta^2 + k_\perp^2} - k_\zeta$. The step $\Delta$ is therefore set by accuracy requirements associated with laser diffraction, plasma and betatron evolution, and, for an undulator, the undulator period.

\subsection{Numerical Cherenkov radiation}\label{sec:cherenkovtheory}

Equation~\eqref{eq:disp} also shows a defect: at the $\zeta$-Nyquist wavenumber, $\theta = \pi$, $\omega = 0$ for every $k_\perp$. This odd--even mode is stationary in the window, like a relativistic bunch, so the macroparticle noise of the bunch current drives it resonantly; its group velocity $\dd\zeta/\dd s = -h^2k_\perp^2/8$ carries it slowly backwards. The result is a numerical Cherenkov radiation~\cite{Godfrey1974} behind the bunch, with $E_x = B_y$ (pure $R$) and a sign change from slice to slice. It grows with the propagation distance and the shot noise, and a smaller $\Delta$ does not help. A $(1,2,1)/4$ filter along $\zeta$ applied to the current of the beam particles, whose response $\cos^2(\theta/2)$ vanishes at $\theta = \pi$, removes the source. The filter acts in index space, so it removes the mode at any local spacing of a user-defined nonuniform grid. The plasma current is not filtered, so the wake is unchanged (Sec.~\ref{sec:cherenkov}).

\subsection{Regularization of the bubble closure}\label{sec:smooth}

The continuum relation of the reduced system in a cold plasma, $\omega = 2\kappa\omega_p^2/(4\kappa^2 - \omega_p^2)$~\cite{GEMPIC}, has a resonance at $\kappa = \omega_p/2$, where $\omega \to \infty$; it is the remnant of the backward branch, which the reduced system does not describe. The trapezoidal rule in $s$ maps such modes onto modes with $\omega\Delta \to \pi$, which change sign from step to step and are not damped. In a uniform plasma they lie at wavelengths longer than the wake period and are not excited. Where the local density is very high, however, $\omega_p$ moves the resonance to resolved scales, and the small, step-dependent jumps of the current that every predictor--corrector and every switch of a particle between levels produce can excite it. In a cold plasma this happens at the back of the bubble (Sec.~\ref{sec:closure}).

As in QUARZ~\cite{QUARZ}, we do not damp the modes but remove the singularity that excites them. After the deposit of every half-cell, the currents and densities of the streaming plasma (electrons, ions and freed electrons) are smoothed radially, per mode,
\begin{align}
\bm j_\perp &\leftarrow (1 - a^2\nabla^2_\mathrm{vec})^{-1}\bm j_\perp, \nonumber\\
(\rho, j_z) &\leftarrow (1 - a^2\nabla^2_m)^{-1}(\rho, j_z), \label{eq:smooth}
\end{align}
which gives every ring of charge a finite width $a$. In contrast to QUARZ, which filters every Cartesian component with the scalar operator, $\nabla^2_\mathrm{vec} = \mathrm{grad}\,\mathrm{div} - \mathrm{rot}\,\mathrm{curl}$ is built from the operators of Sec.~\ref{sec:disc}: $\mathrm{grad}$ is $(G, im/c)$, the curl is CE, $\mathrm{rot}\,b = (im b/\rho, -g b)$, and $\nabla^2_m = \mathrm{div}\,\mathrm{grad}$ is the Laplacian of the Poisson problem. Because $\mathrm{div}\,\mathrm{rot} = 0$ holds exactly on the staggered grid (with $\mathrm{curl} = 0$ on the axis for $m \ge 1$), $\mathrm{div}\,\nabla^2_\mathrm{vec} = \nabla^2_m\,\mathrm{div}$. The filter therefore commutes with the discrete divergence, and the smoothed currents conserve the smoothed charge. The vector system is pentadiagonal ($j_r$ and $j_\varphi$ interleaved) and the scalar one tridiagonal; both are factorized once per mode, and the filter costs less than 1\,\% of the run time. The beam particles and the fields acting on the particles are not smoothed. The length $a$ can be chosen separately for the wake ($m = 0$) and for the laser-driven currents ($m \ge 1$).

The smoothing length is a numerical parameter. It stands for the finite width that the thermal spread and deviations from axial symmetry give the returning electron sheath in a real plasma, but it is not derived from them, so every result that may depend on it has to be checked. Two conditions bound it. From below, the fields behind the closure must converge with the step $\Delta$. From above, $a$ must be small compared with the scales of the physics, since it changes the model by $O(a/\ell)$: for the wake ($m = 0$) $\ell$ is the bubble radius $R_b \approx 2\sqrt{a_0}\,c/\omega_p$, for the laser-driven currents ($m \ge 1$) the laser wavelength and waist. We take the smallest $a$ that meets the first condition and vary it by a factor of two in both directions; Sec.~\ref{sec:sens} summarizes the result.

%---------------------------------------------------------------------------
\section{Implementation}\label{sec:impl}

A scheme becomes a tool only when it runs on the machines where production runs are done and when its results can be trusted and reused. GEM-PIC-RZ is written in C++17 without external dependencies apart from optional MPI and HDF5 libraries. The input is a structured deck in physical units, with profiles of the plasma density, the laser, the bunch current and an undulator taper given as expressions; unknown keys, wrong units and inconsistent settings are reported with file and line. Bunches can also be read from openPMD files of other codes. External elements (quadrupoles, solenoids, active plasma lenses, chicanes, planar and helical undulators) act on the particles only.

\paragraph*{Parallelism.} A parallel engine executes the serial step as kernels behind a portability layer. The code compiles for OpenMP and NVIDIA GPUs and reproduces the serial result bit for bit with OpenMP and to $10^{-9}$ in the emulated GPU paths. Within one step the march is sequential in $\zeta$, so the window is split into blocks of slices, one per MPI rank, and the ranks form a pipeline in the steps~\cite{Feng2009}: while rank $q$ computes step $k$, rank $q+1$, ahead of it in $\zeta$, already computes step $k+1$. The efficiency is $N_s/(N_s + P - 1)$ for $N_s$ steps and $P$ ranks; MPI results agree with the serial run to rounding.

\paragraph*{GPU.} On the GPU each azimuthal mode of the field solve is one thread block. The radial operators are applied by all its threads, and the two tridiagonal systems per slice are solved with the Thomas/PCR partition method~\cite{Laszlo2016}: the rows are cut into chunks of at least eight, each thread eliminates the interior of its chunk, and the reduced system of the chunk ends is solved by parallel cyclic reduction. The matrices change only with the step $\Delta$ and the half-cell lengths, so all matrix-dependent coefficients are computed once on the host, and a solve only processes the right-hand side. The number of chunks depends on the matrix size alone, so the result does not depend on the number of threads; it agrees with the Thomas algorithm to $10^{-12}$. On one NVIDIA GH200 of JUPITER all regression and physics tests pass, and the LWFA benchmark of Sec.~\ref{sec:fbpic} reproduces the deviations from FBPIC of the CPU run to all printed digits. For that benchmark with 16 times more plasma particles (about $4\times10^4$ per slice, 200 steps), a step takes 0.37\,s on one GH200. Half of this time is spent in the push and deposition of the plasma particles and 22\,\% in the field solve; the partition method halved the latter compared with a one-thread Thomas solve (31\,s $\to$ 17\,s for the run). Runs with few particles per slice, such as the undulator cases, are limited by the latency of the about ten kernel launches per slice.

\paragraph*{Output and tests.} Checkpoints hold the complete state, and a restart is exact on any number of ranks and back-ends. openPMD files~\cite{openPMD} hold the fields interpolated to a uniform grid and the exact data on the native nonuniform nodes, in the form we have proposed, together with QUARZ~\cite{QUARZ}, for the openPMD 2.0 standard~\cite{openPMDPR}. A test suite of 25 tests, with about 230 unit checks, runs in minutes and includes all tests of Sec.~\ref{sec:valid}.

Like QUARZ~\cite{QUARZ}, GEM-PIC-RZ was developed using a \emph{zero-touch} workflow: the AI system Claude~\cite{Claude} generated the code, tests, benchmark scripts, and documentation, while the author directed the work and did not write or edit the source code. The author's contributions included the GEM-PIC formulation~\cite{GEMPIC}, the selection of numerical methods and benchmark problems, physical conventions (including the Galilean transformation used for the particles), the verification criteria, and interpretation of the results. The development process required each feature to pass tests against independent analytic or numerical references, with the full test suite rerun after each change. Several numerical errors discussed in this paper were identified when tests failed against their references. The public repository~\cite{GEMPICRZ} records the AI session associated with each commit. This disclosure describes the software-development process; the author remains responsible for the manuscript and all scientific claims.

%---------------------------------------------------------------------------
\section{Verification}\label{sec:valid}

Before comparing with other codes, the building blocks have to be shown to be correct against references that do not depend on any simulation code. Table~\ref{tab:tests} lists these tests: the self-field of a relativistic bunch against Gauss's law on uniform and stretched grids, laser diffraction in vacuum, the collapse of ions in a dense bunch, and field ionization against rate equations.

\begin{table}[htb]
\caption{Verification tests (all in the test suite).}\label{tab:tests}
\begin{ruledtabular}
\begin{tabular}{p{0.52\columnwidth}p{0.4\columnwidth}}
test & result \\ \hline
Field of an ultra-relativistic bunch in vacuum vs Gauss's law (uniform, stretched and user-defined nonuniform radial grids) & $E_r$ to $5\times10^{-4}$; noise $E_z < 2\times10^{-3}E_r$ \\
Same on a user-defined nonuniform $\zeta$ grid (2.5--8\,$k_0^{-1}$ across the bunch) & $2\times10^{-3}$ (shot-noise level) \\
$\gamma = 100$ bunch in its own field, 8\,cm & energy kept to $2\times10^{-5}$ \\
Laser focusing in vacuum ($w_0 = 5\,\mu$m, $z_R = 98\,\mu$m, 0.4\,mm) & $a_0(z)$ within 3\,\% \\
Ion collapse in a dense bunch ($n_b = 100 n_0$, H)~\cite{Rosenzweig2005} vs test-ion model & on-axis $n_i$ 1--3\,\%; D/H delay 1.418 ($\sqrt2$) \\
Field ionization of H vs $1 - \exp(-\int w\,\dd t)$ & 0.02294 vs 0.02288 \\
N$^{5+}\to$N$^{6+}\to$N$^{7+}$ ($a_0 = 2.2$) vs rate equations & 0.9357/0.0600/0.0044 vs 0.9367/0.0593/0.0041 \\
ADK rate vs Landau rate (H-like ions) & $10^{-12}$ \\
Immobile limit of ions ($M = 10^{15}$\,u) & identical fields \\
\end{tabular}
\end{ruledtabular}
\end{table}

\subsection{User-defined nonuniform $\zeta$ and $s$ grids}

The user-defined nonuniform grids are meant to save computing time without loss of accuracy. The test is a beam-driven wake with two short bunches far apart, the configuration of a plasma wakefield accelerator with a driver and a witness: a plasma of $0.002\,n_c$, a driver of $\sigma_z = 1.5\,k_0^{-1}$ and a witness $130\,k_0^{-1}$ behind (Table~\ref{tab:zgrid}). A uniform coarse grid ($h = 2$) is 16\,\% off at the driver; a user-defined nonuniform grid that is fine ($h = 0.25$) only at the two bunches reproduces the fine uniform grid to $10^{-3}$ with a quarter of the slices and 3.6 times faster. In the laser wakefield case of Sec.~\ref{sec:fbpic}, a step in $s$ that is small only in the density up-ramp gives the witness energy within 0.4\,\% of a run with uniformly small steps, with 44 instead of 80 steps; uniform large steps are 1.4\,\% off.

\begin{table}[htb]
\caption{Beam-driven wake on uniform and user-defined nonuniform $\zeta$ grids. $E_z$ errors relative to $\max|E_z|$, witness energy change over 50 steps.}\label{tab:zgrid}
\begin{ruledtabular}
\begin{tabular}{lccc}
$\zeta$ grid & slices & $E_z$ error\footnotemark[1] & $\Delta E$ error \\ \hline
uniform $h = 0.25$ (ref.) & 2800 & -- & -- \\
uniform $h = 2$ & 350 & 16\,\% / 6.6\,\% & 1.3\,\% \\
nonuniform 0.25 / 2 & 734 & 0.11\,\% / 0.84\,\% & 0.06\,\% \\
\end{tabular}
\footnotetext[1]{At the driver / at the witness.}
\end{ruledtabular}
\end{table}

\subsection{Numerical Cherenkov radiation}\label{sec:cherenkov}

Relativistic bunches that propagate over long distances, in a plasma wakefield accelerator or an undulator, are where numerical Cherenkov radiation would spoil a run. A bunch of 1\,pC at $\gamma = 1000$ ($\sigma_r = 2$, $\sigma_z = 4\,k_0^{-1}$, 16 slices) propagates 1000 steps of $\Delta = 20\,k_0^{-1}$ in vacuum. Without the filter of Sec.~\ref{sec:cherenkovtheory} the $\zeta$-Nyquist part of the field energy grows to $8.6\times10^{-2}$ of the total; with the filter it stays at $1.3\times10^{-4}$, the floor of the estimator. In a plasma wake ($0.002\,n_c$, 0.2\,pC driver) the Nyquist fraction grows from $1.3$ to $1.9\times10^{-2}$ between 200 and 400 steps without the filter and stays at $8\times10^{-5}$ with it, while the bunch parameters agree to four digits and the wake to 0.35\,\%.

%---------------------------------------------------------------------------
\section{Laser-plasma acceleration: comparison with full PIC}\label{sec:lpa}

We compare GEM-PIC-RZ with the quasi-cylindrical spectral full PIC code FBPIC~\cite{Lehe2016} for key processes in laser wakefield acceleration: nonlinear wake formation in the bubble regime, including bubble closure; laser evolution over multiple Rayleigh lengths; and electron injection by ionization and by a density down-ramp. In all cases FBPIC uses two azimuthal modes and a cell size of $\lambda_0/24$ to $\lambda_0/32$ along $z$. For the comparisons it matters that a level $s$ of GEM-PIC-RZ is the plane $z = s$ and its slices are the times at which the laser and the wake pass that plane (GT2), while an FBPIC output is a snapshot at one time. Wakes are compared on the axis as functions of $\zeta$, and trapped electrons at the same lab position of the bunch. All decks and analysis scripts are in the repository.

\subsection{The wake in the bubble regime}\label{sec:fbpic}

\begin{figure}[htb]
\includegraphics[width=\columnwidth]{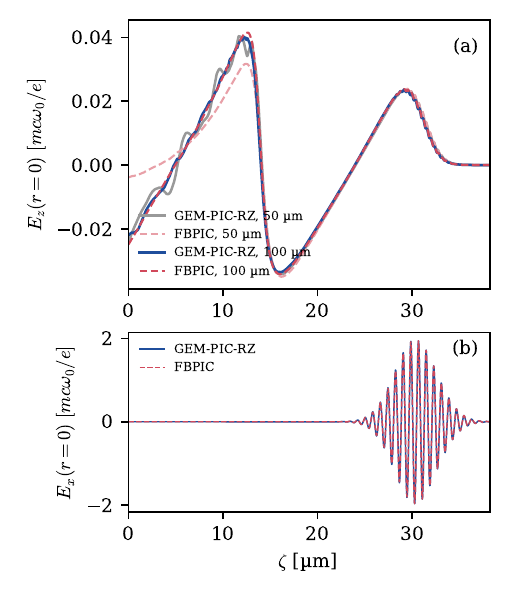}
\caption{Laser wakefield in the bubble regime ($a_0 = 2$, $0.002\,n_c$): (a) on-axis $E_z$ after 50 and 100\,$\mu$m of propagation, GEM-PIC-RZ (solid) and FBPIC (dashed); (b) the laser field $E_x$ on the axis at 100\,$\mu$m. The laser moves to the right.}\label{fig:lwfa}
\end{figure}

Many laser wakefield accelerators operate in the bubble (blowout) regime~\cite{Pukhov2002,Lu2007}. The laser ponderomotive force expels plasma electrons, leaving an ion cavity behind the pulse. The cavity fields accelerate and focus injected electrons, while the cavity shape and length determine the energy gain. The configuration is a typical one for a 100-TW-class laser: $\lambda_0 = 0.8\,\mu$m, $a_0 = 2$, $w_0 = 8.02\,\mu$m and a field duration of 10.6\,fs, linearly polarized and starting at focus, in a plasma of $3.48\times10^{18}$\,cm$^{-3}$ ($0.002\,n_c$) after a 20\,$\mu$m linear ramp, with immobile ions. GEM-PIC-RZ uses 100 radial cells, $0.3\,k_0^{-1}$ on the axis, stretched to $3\,k_0^{-1}$ outside, $h = 0.25\,k_0^{-1}$, $\Delta = 2.5\,\mu$m (100 times the FBPIC time step $c\,\delta t = \lambda_0/32$), and the source smoothing $a = 2\,k_0^{-1}$ of Sec.~\ref{sec:smooth}.

Figure~\ref{fig:lwfa} shows the on-axis fields. After 100\,$\mu$m the first wake period agrees to 1.9\,\% rms of the peak field, the minimum of $E_z$ is $-0.0335$ vs $-0.0342\,m_ec\omega_0/e$, and the laser peak field 1.969 vs 1.954. In the up-ramp at 50\,$\mu$m the codes differ by representation, not by physics: GEM-PIC-RZ shows the wake as it passes one plane, with the density of that plane, while FBPIC shows the wake behind the laser at smaller $z$, where the density is lower. The difference disappears where the density is constant over a window length. On one CPU core GEM-PIC-RZ took 1.7\,min, FBPIC about 1.5\,h.

\subsection{Bubble closure}\label{sec:closure}

\begin{figure}[htb]
\includegraphics[width=\columnwidth]{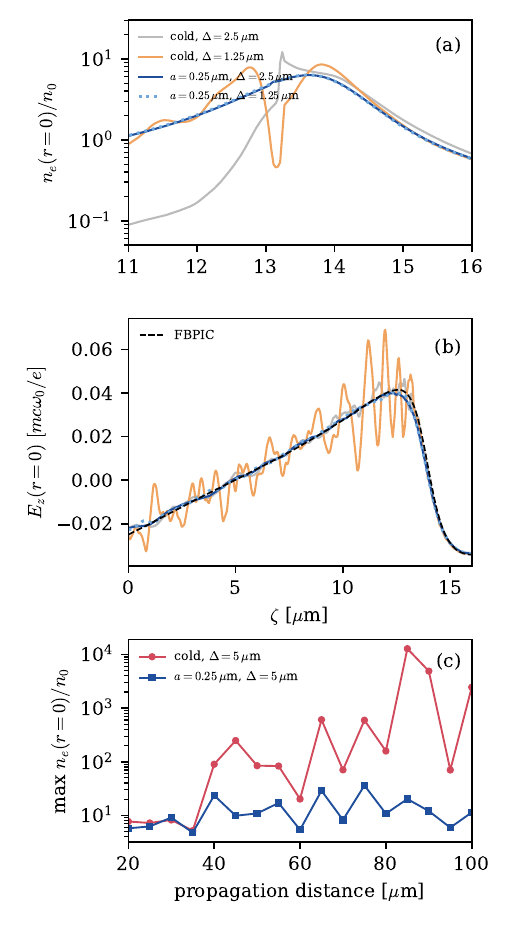}
\caption{Back of the first bubble in the case of Fig.~\ref{fig:lwfa} at 100\,$\mu$m: (a) on-axis electron density, (b) on-axis $E_z$, cold plasma (grey, orange) and with source smoothing $a = 0.25\,\mu$m (blue), for $\Delta = 2.5$ and $1.25\,\mu$m, and FBPIC (dashed); (c) peak on-axis density at the closure against the propagation distance for $\Delta = 5\,\mu$m, cold (red) and smoothed (blue).}\label{fig:closure}
\end{figure}

At the back of the bubble the expelled electrons return to the axis. There the accelerating field is strongest, there the electrons are injected, and there the witness bunch of a two-bunch scheme sits, so the fields in this region decide the quality of the accelerated beam. In a cold plasma with an axisymmetric driver the returning electron trajectories cross on the axis and form a caustic~\cite{Lotov2003}: the density is singular, and on a fine radial grid it grows without bound as the cells are refined. Uniform grids of full PIC codes cut the singularity off implicitly over one cell; the fine axis cells of GEM-PIC-RZ expose it. The singular density moves the resonance of the reduced equations (Sec.~\ref{sec:smooth}) to resolved scales, and the step-dependent perturbations of the current excite it.

Figure~\ref{fig:closure} shows the result for the wake of Fig.~\ref{fig:lwfa}. In the cold plasma the density at the closure is a narrow spike whose height and shape change with $\Delta$ [Fig.~\ref{fig:closure}(a)], and the wake behind it carries oscillations on the scale of $\lambda_0$ that do not converge with $\Delta$: the rms deviation from FBPIC is $1.8\times10^{-3}\,m_ec\omega_0/e$ for $\Delta = 2.5\,\mu$m and $9.8\times10^{-3}$ for $\Delta = 1.25\,\mu$m [Fig.~\ref{fig:closure}(b)]. With $\Delta = 5\,\mu$m the closure density alternates from step to step between about $70$ and $10^4\,n_0$, the signature of the period-two modes, and the peak field reaches $0.86\,m_ec\omega_0/e$, twenty times the wake [Fig.~\ref{fig:closure}(c)].

The spike is unphysical: real plasmas are warm, real drivers are not perfectly cylindrical, and the spike carries no energy. The smoothing of Eq.~\eqref{eq:smooth} with $a = 2\,k_0^{-1}$ (0.25\,$\mu$m, $0.09\,c/\omega_p$, about the axis cell) replaces it by a peak of finite width. The oscillations disappear: the rms deviation behind the bubble is $0.7\times10^{-3}\,m_ec\omega_0/e$ for both $\Delta = 2.5$ and $1.25\,\mu$m, and the two runs agree with each other to the same level, and with $\Delta = 5\,\mu$m the closure density stays bounded at 5--40\,$n_0$. The price is a model change of order $a$ at the bubble back: the peak of $E_z$ there changes from 0.047 (the spike) to 0.040 (FBPIC: 0.042), and the rms agreement of the first wake period from 1.6 to 1.9\,\%. The criterion is that the fields behind the spike converge, not the height of the spike. In a denser plasma the closure is stronger and the wake needs a larger $a$ (Sec.~\ref{sec:laser}). Two alternatives that also suppress the oscillations, an off-centred update of $R$ in $s$ and keeping electrons in their level for several steps, are available as options; the first damps the laser, the second changes the trapping (Sec.~\ref{sec:downramp}).

\subsection{Laser evolution: diffraction, depletion and redshift}\label{sec:laser}

\begin{figure}[htb]
\includegraphics[width=\columnwidth]{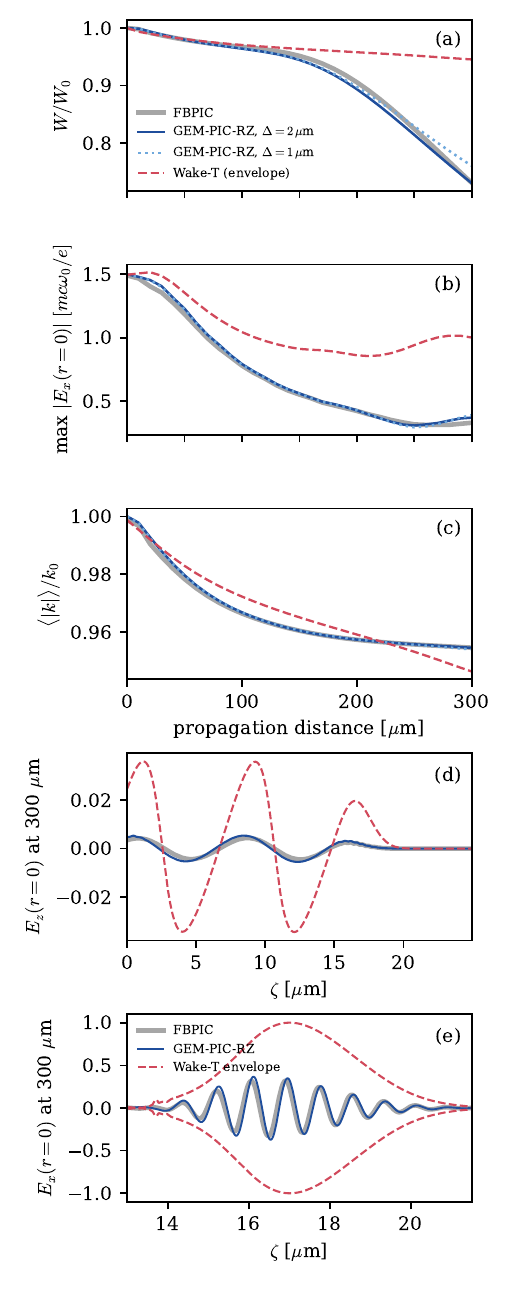}
\caption{Evolution of a 6.8\,fs laser pulse ($a_0 = 1.5$) in a plasma of $0.01\,n_c$ over 300\,$\mu$m (six Rayleigh lengths): (a) laser energy, (b) peak of $|E_x|$ on the axis, (c) spectral centroid of the laser fields; (d) on-axis $E_z$ and (e) on-axis laser field $E_x$ at 300\,$\mu$m (for Wake-T $\pm$ the envelope). FBPIC (grey), GEM-PIC-RZ with $\Delta = 2$ and $1\,\mu$m (blue) and the envelope code Wake-T (red, dashed).}\label{fig:laser}
\end{figure}

The energy gain of a laser wakefield stage is limited by the depletion of the laser, which transfers its energy to the wake and is shifted to lower frequencies in the process~\cite{Decker1996,Lu2007}. Following the laser over the full stage is the task for which GEM-PIC was designed, and it is where envelope models, which assume a slowly varying envelope, are most strained: for few-cycle pulses in dense plasmas, as used with kHz lasers, the pulse depletes within a few Rayleigh lengths and its spectrum changes by several percent. The configuration is such a case: $a_0 = 1.5$, $w_0 = 3.56\,\mu$m ($P/P_c = 0.56$), a field duration of 6.8\,fs (2.5 laser cycles), starting at focus in a plasma of $0.01\,n_c = 1.74\times10^{19}$\,cm$^{-3}$, followed over 300\,$\mu$m. We compare with FBPIC and with the envelope code Wake-T~\cite{FerranPousa2019}, which combines a laser envelope solver with a quasi-static plasma model.

Over 300\,$\mu$m the pulse diffracts, loses a quarter of its energy to the wake and is redshifted by 4.5\,\% (Fig.~\ref{fig:laser}). GEM-PIC-RZ ($h = 0.2\,k_0^{-1}$, $\Delta = 2\,\mu$m) agrees with FBPIC over the whole distance: the energy to $0.013\,W_0$, the redshift to $0.002\,k_0$, the peak field to 5\,\% (13\,\% at the end, where the pulse is weak and spread out), and the first wake period on the axis to 6\,\% rms of its peak. The carrier itself agrees as well [Fig.~\ref{fig:laser}(e)]: at 300\,$\mu$m, after a plasma phase slip of $k_0 z\,n_e/2n_c \approx 12$\,rad, the on-axis field of GEM-PIC-RZ is shifted against FBPIC by 0.4\,rad ($\lambda/16$), and its peak is 12\,\% higher. Further behind, where the wake of the depleted pulse is weak, phase differences raise the rms deviation to about 20\,\%. With $\Delta = 1\,\mu$m the results change by less than $0.03\,W_0$. In the envelope model the pulse stays focused, drives a much stronger wake and loses only 5.5\,\% of its energy instead of 27\,\%; the Wake-T results do not change with resolution. GEM-PIC-RZ took 6.7\,min on one core, FBPIC 3.7\,h on two cores.

At this density the closure of Sec.~\ref{sec:closure} is strong enough to make the cold-plasma run unstable several wake periods behind the laser, where $E_z$ exceeds the FBPIC value more than tenfold; the laser itself is hardly affected ($W = 0.740\,W_0$). The wake is stable for an $m = 0$ smoothing length of at least $2\,k_0^{-1}$ ($0.2\,c/\omega_p$); with $1\,k_0^{-1}$ it still deviates by 80\,\% rms at 300\,$\mu$m, and lengths of 2, 3 and $4\,k_0^{-1}$ give the same laser and the same wake. The laser-driven $m = 1$ currents need much less: with $a = 0.5$, 1 and $2\,k_0^{-1}$ the laser energy at 300\,$\mu$m is 0.726, 0.731 and $0.747\,W_0$, while $3\,k_0^{-1}$, a fifth of the laser wavelength, gives $0.773\,W_0$, 6\,\% above FBPIC. The run of Fig.~\ref{fig:laser} uses $a = 1\,k_0^{-1}$ for $m \ge 1$ and $3\,k_0^{-1}$ for $m = 0$.

\subsection{Ionization injection}\label{sec:ioninj}

\begin{figure}[htb]
\includegraphics[width=\columnwidth]{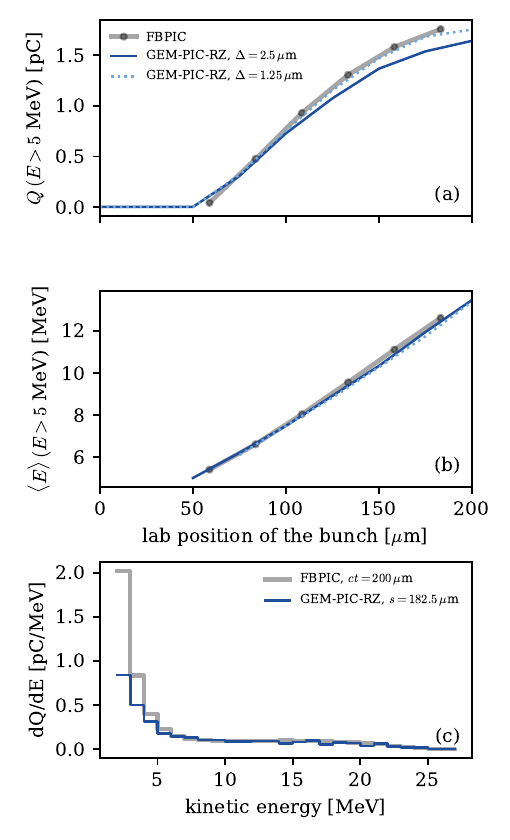}
\caption{Ionization injection from a nitrogen dopant ($a_0 = 2.3$, $0.002\,n_c$, 1\,\% N$^{5+}$): (a) charge and (b) mean energy of the injected electrons above 5\,MeV against the lab position of the bunch, (c) their spectrum at $z = 183\,\mu$m. FBPIC (grey) and GEM-PIC-RZ with $\Delta = 2.5$ and $1.25\,\mu$m (blue).}\label{fig:ioninj}
\end{figure}

Ionization injection~\cite{Pak2010,McGuffey2010} is one of the most widely used ways to load electrons into a laser wakefield: a small admixture of a high-$Z$ gas releases its inner-shell electrons near the peak of the laser field, inside the bubble, where they are trapped. The process depends on the instantaneous field of the carrier, which GEM-PIC-RZ resolves and an envelope code has to model with averaged rates, and it tests field ionization, trapping and acceleration together. The configuration is the case of Sec.~\ref{sec:fbpic} with $a_0 = 2.3$, still without self-injection, and N$^{5+}$ ions at 1\,\% of the electron density, which both codes ionize to N$^{6+}$ and N$^{7+}$ by the ADK rate.

The ionized fraction of the dopant agrees (3.1\,\% N$^{6+}$ and 0.3\,\% N$^{7+}$ behind the laser), and so does the injected bunch (Fig.~\ref{fig:ioninj}). At $z = 183\,\mu$m its charge above 5\,MeV is 1.58\,pC against 1.76\,pC in FBPIC, its mean energy 12.4 against 12.6\,MeV and its rms energy spread 5.2 against 5.4\,MeV, and the spectra agree up to the highest energy. Halving $\Delta$ changes the charge by 7\,\%. Below about 4\,MeV FBPIC holds more electrons: sheath electrons that are not trapped and that GEM-PIC-RZ streams through the window. The injected charge, unlike its energy, depends on the smoothing length, because the trapping is decided at the closure where the smoothing acts (Sec.~\ref{sec:sens}). GEM-PIC-RZ took 8.5\,min on one core, FBPIC 3.4\,h on two.

\subsection{Density down-ramp injection: an open problem}\label{sec:downramp}

\begin{table}[htb]
\caption{Density down-ramp injection: charge (pC) of the trapped electrons above 2 and 5\,MeV. FBPIC at $ct = 150$ and $200\,\mu$m (bunch at $z = 136$ and $179\,\mu$m); GEM-PIC-RZ at the levels $s = 150$ and $200\,\mu$m. The last row uses $\Delta = 0.125\,\mu$m in the ramp and $1.25\,\mu$m elsewhere.}\label{tab:downramp}
\begin{ruledtabular}
\begin{tabular}{lcccc}
 & \multicolumn{2}{c}{$ct$ or $s = 150\,\mu$m} & \multicolumn{2}{c}{$ct$ or $s = 200\,\mu$m} \\
 & $>2$\,MeV & $>5$\,MeV & $>2$\,MeV & $>5$\,MeV \\
\hline
FBPIC & 51.0 & 14.6 & 51.1 & 30.7 \\
$\Delta = 2.5\,\mu$m & 9.9 & 8.9 & 8.5 & 8.4 \\
$\Delta = 1.25\,\mu$m & 20.9 & 17.1 & 19.6 & 18.8 \\
$\Delta = 0.5\,\mu$m & 12.2 & 9.9 & -- & -- \\
$\Delta = 0.25\,\mu$m & 14.2 & 13.7 & -- & -- \\
ramp refined & -- & -- & 27.6 & 12.6 \\
\end{tabular}
\end{ruledtabular}
\end{table}

In a density down-ramp the plasma wavelength grows, the back of the bubble slows down, and background electrons that it overtakes are trapped~\cite{Bulanov1998,Geddes2008,Buck2013}. Sharp ramps, made with shocks or blades in a gas jet, give bunches with small energy spread and are used in many experiments. We report this benchmark as a caution, because GEM-PIC-RZ does not yet pass it. The laser of Sec.~\ref{sec:fbpic}, now with $a_0 = 1.8$, crosses a plateau of $2n_0$ ($n_0 = 0.002\,n_c$) that drops linearly to $n_0$ between $z = 60$ and $70\,\mu$m. Behind the ramp the bubble lengthens by about 30\,\% within 10\,$\mu$m, and its back moves forward through the window at about $0.5\,c$.

The codes agree qualitatively: both start trapping at the ramp (on a constant plateau of $2n_0$ GEM-PIC-RZ traps no electrons), and the wake on the plateau before the ramp agrees to $2\times10^{-3}\,m_ec\omega_0/e$ rms. They disagree quantitatively (Table~\ref{tab:downramp}). FBPIC traps 51\,pC above 2\,MeV; GEM-PIC-RZ traps between 9 and 28\,pC, and the charge depends on $\Delta$ without converging. The difference persists with a finer step inside the ramp alone, with a higher velocity threshold for the beam list and with stronger source smoothing; keeping newly trapped electrons on the streaming list for several steps suppresses the trapping almost entirely.

We hypothesize that the discrepancy is related to the use of GT2 for the particle advance. In GT2 a level is a plane $z = s$, and an electron is trapped when its lab position reaches the next plane inside the window; the decision is made in the fields of one plane, while the back of the bubble, which does the trapping, moves through the window by a large fraction of its length between two planes. In GT1 a level is an instant of time, as in FBPIC, and the moving back of the bubble is followed in time; this frame should describe down-ramp injection. GT1 has its own difficulty, however: plasma particles that reach the next slow-time level inside the window stop being pushed there, and this imprints stripes on the fields. A particle scheme that combines the two will be the subject of a follow-up study. Until then, GEM-PIC-RZ results for self-injection in sharp density transitions should not be trusted quantitatively, whereas ionization injection (Sec.~\ref{sec:ioninj}), where the electrons are born inside a slowly evolving bubble, is reproduced.

\subsection{Robustness to the smoothing length and the step}\label{sec:sens}

\begin{table*}[htb]
\caption{Sensitivity of the benchmark results to the source-smoothing length $a$ (in $k_0^{-1}$) and the step $\Delta$. Pairs give the result for the step of the benchmark and for half of it ($\Delta = 2.5/1.25\,\mu$m; laser: $2/1\,\mu$m). Laser evolution: $a$ is the length for $m \ge 1$, with $3\,k_0^{-1}$ for $m = 0$ (cold: both zero); with $a = 3$ for $m \ge 1$, $W/W_0 = 0.773$. Ionization injection: bunch at $z = 175\,\mu$m, FBPIC interpolated to it. Fields in $m_ec\omega_0/e$, deviations from FBPIC rms, in \% of the peak or in $10^{-3}\,m_ec\omega_0/e$. Bold: the settings of the benchmark figures.}\label{tab:sens}
\begin{ruledtabular}
\footnotesize
\begin{tabular}{llccccc}
benchmark & quantity & FBPIC & $a = 0$ & $a = 1$ & $a = 2$ & $a = 4$ \\ \hline
wake, 100\,$\mu$m & rms dev., first period [\%] & -- & 1.6 / 2.0 & 1.8 / 2.0 & \textbf{1.9} / 2.1 & 3.4 / 3.6 \\
 & rms dev., behind bubble [$10^{-3}$] & -- & 1.8 / 9.7 & 1.1 / 1.8 & \textbf{0.70} / 0.72 & 1.5 / 1.9 \\
 & $E_z$ at bubble back & 0.042 & 0.047 / 0.069 & 0.042 / 0.042 & \textbf{0.040} / 0.041 & 0.036 / 0.035 \\
laser, 300\,$\mu$m & $W/W_0$ & 0.731 & 0.740 & \textbf{0.731} / 0.761 & 0.747 & -- \\
 & $\langle|k|\rangle/k_0$ & 0.9547 & 0.9549 & \textbf{0.9546} / 0.9541 & 0.9535 & -- \\
 & rms dev., wake [\%] & -- & 1300 & \textbf{19} / 24 & 16 & -- \\
ionization inj. & $Q(E > 5\,\mathrm{MeV})$ [pC] & 1.70 & 1.99 & 1.84 / 1.95 & \textbf{1.54} / 1.68 & 0.96 \\
 & $\langle E\rangle$ [MeV] & 12.1 & 12.3 & 12.1 / 12.0 & \textbf{11.9} / 11.7 & 11.0 \\
down-ramp inj. & $Q(E > 2\,\mathrm{MeV})$, 150\,$\mu$m [pC] & 51.0 & 18.8 / 21.9 & 14.4 & \textbf{9.9} / 20.9 & 10.1 / 17.0 \\
\end{tabular}
\end{ruledtabular}
\end{table*}

The source smoothing is a numerical model parameter, and a result is only as good as its independence of it. We therefore repeated every benchmark with smoothing lengths from 0 to $4\,k_0^{-1}$ and with half the step (Table~\ref{tab:sens}; scripts and runs in the repository). The conclusions fall into three classes.

\emph{Robust.} The wake up to the back of the first bubble agrees with FBPIC to 1.6--2.1\,\% for all $a \le 2\,k_0^{-1}$ and both steps. The laser energy and redshift after six Rayleigh lengths agree to within 0.016\,$W_0$ and 0.0012\,$k_0$ for all $m \ge 1$ lengths up to $2\,k_0^{-1}$ and all $m = 0$ lengths that keep the wake stable, and the half step changes the energy by 0.03\,$W_0$; even the unstable cold run gives the laser correctly. The energy of the ionization-injected electrons (11.7--12.3\,MeV against 12.1\,MeV) does not depend on $a \le 2\,k_0^{-1}$ or $\Delta$. These conclusions do not rest on the choice of $a$.

\emph{Bounded by the choice of $a$.} The wake behind the closure needs a minimum smoothing length. Below it the result does not converge with $\Delta$ (cold: 1.8 against $9.7\times10^{-3}$; $a = 1\,k_0^{-1}$: 1.1 against $1.8\times10^{-3}$); at $a = 2\,k_0^{-1}$ it converges and agrees best with FBPIC, at $4\,k_0^{-1}$ it converges with a larger model error, and the field at the bubble back drops to 0.036 (FBPIC 0.042). In the denser plasma of Sec.~\ref{sec:laser} the minimum is $2\,k_0^{-1}$ for $m = 0$, about $0.2\,c/\omega_p$ against $0.09\,c/\omega_p$ at $0.002\,n_c$, and the laser-driven currents tolerate up to $2\,k_0^{-1}$. The lengths used in the figures are the smallest that converge, and the results quoted for them hold within the stated agreement for a factor of two around them.

\emph{Sensitive.} The injected charge depends on $a$ because trapping is a threshold process decided at the closure, where the smoothing acts. Above 5\,MeV it ranges from 1.99 to 1.54\,pC for $a = 0$ to $2\,k_0^{-1}$, with FBPIC (1.70\,pC) inside this range for both steps, and drops to 0.96\,pC at $a = 4\,k_0^{-1}$. The agreement of Sec.~\ref{sec:ioninj} in charge is therefore to be read with a model uncertainty of about $\pm15\,\%$, larger than the 7\,\% from the step; the energy and the spectrum are robust. For the down-ramp the table shows the converse: for every $a$ and $\Delta$ GEM-PIC-RZ traps 10--22\,pC against 51\,pC, so the discrepancy of Sec.~\ref{sec:downramp} is not caused by the regularization.

%---------------------------------------------------------------------------
\section{Undulator radiation and free-electron lasers}\label{sec:fel}

Because the carrier is resolved, the same equations describe the radiation of the accelerated beam. A beam that wiggles in an undulator radiates through the transverse current in Eqs.~\eqref{eq:Rr}, \eqref{eq:Rp}, and the radiation acts back on the beam; it stays at fixed $\zeta$ while the electrons slip back by $\lambda_r$ per period. Spontaneous and coherent emission and the FEL interaction are therefore contained in the model, with the full space-charge fields of the beam and without the period averaging of dedicated FEL codes~\cite{Reiche1999}. The $\zeta$ grid resolves $\lambda_r$, the step in $s$ the undulator period. First tests are encouraging: the coherent spontaneous emission of a short bunch agrees with the far-field radiation integral to 2--4\,\% in energy, the SASE gain length of a compact FEL with the Ming Xie fit~\cite{Xie1995} (6.3 vs 6.4 periods), and seeded amplifiers at 10\,nm and 0.8\,nm with GENESIS~1.3~\cite{Reiche1999} to 4\,\% and 0.6\,\% in gain length. A combined simulation of plasma acceleration, beam transport, and FEL amplification in a single code will be reported separately.

%---------------------------------------------------------------------------
\section{Conclusions}\label{sec:concl}

GEM-PIC-RZ extends the Galilean electromagnetic PIC scheme to quasi-cylindrical geometry. It resolves the laser carrier while avoiding a Courant condition that ties the propagation step to the cell size. In the laser wakefield benchmarks, its propagation step is about 100 times the FBPIC time step; the practical step size remains limited by accuracy and, in plasma, by the condition $\Delta h \lesssim 4/\omega_p^2$ (Sec.~\ref{sec:dispersion}). Two numerical problems of the scheme were found and solved: a stationary odd--even mode that radiates behind relativistic bunches, removed by a filter of the beam current, and the excitation of spurious modes by the singular bubble closure of a cold plasma, removed by a charge-conserving radial smoothing of the plasma sources.

Against the full PIC code FBPIC, GEM-PIC-RZ reproduces the bubble-regime wake, the depletion and redshift of a few-cycle pulse over six Rayleigh lengths, where an envelope model fails, and ionization injection, in charge, energy and spectrum, at about 2\,\% of the CPU time. This makes it suited for parameter scans and for the design of laser wakefield stages over their full length. These conclusions hold for smoothing lengths within a factor of two around the values used and for half the step; the injected charge is the exception, with a model uncertainty of about 15\,\% from the regularization. Two limits remain. Backward-propagating radiation (Raman backscatter, reflection from overdense plasma) is outside the model, since $\partial_s L$ is dropped. Self-injection in a sharp density down-ramp is not yet reproduced; we relate this to the GT2 frame of the particle advance and will address it, together with the artifacts of the GT1 frame, in a follow-up study. Planned work also includes benchmarks of beam-driven wakes, the combination of the GPU engine with the MPI pipeline, fewer kernel launches per slice, and the FEL applications.

GEM-PIC-RZ is open source~\cite{GEMPICRZ}, with the decks and scripts of all tests in this paper. Its development, like that of QUARZ, shows that a production-grade simulation code can be built by an AI agent under the direction of a physicist, when every feature is held to an independent test.

\begin{acknowledgments}
This work was supported by DFG and BMBF. The author gratefully acknowledges the Gauss Centre for Supercomputing e.V. (gauss-centre.eu) for funding this project (lpqed) by providing computing time through the John von Neumann Institute for Computing (NIC) on the GCS Supercomputer JUPITER at J\"ulich Supercomputing Centre (JSC). The GPU tests have been run by Dr. Thomas C. Wilson.
\end{acknowledgments}

\end{document}